\documentclass{ws-ijmpe}
\usepackage[super,compress]{cite}
\begin{document}

\markboth{M. Foley, N. Sasankan, M. Kusakabe, G. J. Mathews}{Revised Uncertainties in Big Bang Nucleosynthesis}

\catchline{}{}{}{}{}

\title{Revised Uncertainties in Big Bang Nucleosynthesis} 

\author{\footnotesize M. Foley
}

\address{Center for Astrophysics, Department of Physics, University of Notre Dame\\
Notre Dame, IN 46556,
USA
mfoley7@nd.edu}

\author{N. Sasankan}
\address{Center for Astrophysics, Department of Physics, University of Notre Dame\\
Notre Dame, IN 46556,
USA}

\author{M. Kusakabe}
\address{Center for Astrophysics, Department of Physics, University of Notre Dame\\
Notre Dame, IN 46556,
USA}

\author{G. J. Mathews}
\address{Center for Astrophysics, Department of Physics, University of Notre Dame\\
Notre Dame, IN 46556,
USA}

\maketitle

\begin{history}
\received{11 December 2016}
\revised{7 June 2016}
\end{history}

\begin{abstract}
Big Bang Nucleosynthesis (BBN) explores the first few minutes of nuclei formation after the Big Bang.  We present updates that result in new constraints at the 2$\sigma$ level for the abundances of the four primary light nuclides - D,$^3$He,$^4$He, and $^7$Li - in BBN. A modified  standard BBN code was used in a Monte Carlo analysis of the nucleosynthesis uncertainty as a function of  baryon-to-photon ratio. Reaction rates were updated to those of NACRE and  REACLIB, and R-Matrix calculations.    The results are then used to derive a new constraint on  the effective number of neutrinos. 
 \end{abstract}

\ccode{25.35.+c,95.30.-k}
\maketitle


\section{\label{sec:level1}Introduction}

Big Bang Nucleosynthesis (BBN) begins approximately 1 second into the Big Bang and lasts for about 3 minutes. BBN is the only window into the conditions of the early universe before the CMB. During this epoch, the temperature is optimal for the formation of light nuclei, resulting in the synthesis of D,$^3$He,$^4$He, and $^7$Li. 

After the quark-hadron transition, the neutron-to-proton ratio $n/p$ is held in equilibrium through weak interactions with electrons, neutrinos, and their antiparticles. However, as temperature drops, neutrinos decouple from the background and the $n/p$ is no longer in equilibrium. When the temperature has dropped below about $T$ = 0.7 MeV, the $n/p$ freezes at $\approx$1/6. Subsequently, almost all neutrons are incorporated into helium nuclei. The details of this are dependent upon the nuclear reaction rates, the baryon densities, and the cosmic expansion as the nuclear reactions freeze out. The final abundances are therefore sensitive to cosmological parameters as noted in Refs.~[\refcite{Mathews17,Coc17,Nakamura17}] of this volume (see also the recent review of Cyburt, Fields and Olive.\cite{Cyburt16}). 

Light element abundances are governed by nuclear reactions and their freeze out through cosmic expansion. As the universe expands according to the Friedmann equation, the temperature drops, eventually allowing for the formation of nuclei. The cosmic expansion rate $H=\dot{a}/a$ is defined with the scale factor of the universe $a(t)$. The Friedmann equation is given by:

\begin{equation}
H^{2}=\frac {8}{3}\pi G \rho + \frac {\Lambda}{3} - \frac {k}{a^2}
\end{equation}

Here $\rho$ is the total energy density, $G$ is the gravitational constant, $k$ is the curvature and $\Lambda$ is the cosmological constant. During BBN, a flat universe ($k = 0$) can be assumed since the actual curvature term is small during the radiation-dominated epoch. 
The nuclear reaction rates are given by the Maxwellian-averaged cross section:

\begin{equation}
<\sigma v>=\frac{(8/\pi)^{1/2}}{\mu^{1/2} (k_B T)^{3/2}} \int _{0}^{\infty}E \sigma (E) \mbox{exp}[-E/(k_B T)]dE
\end{equation}

Here $\sigma (E)$ is the cross section at center of mass energy $E = \mu v^{2}/2$ where $v$ is the relative velocity and $\mu = m_A m_B/(m_A + m_B)$ is the reduced mass of nuclei $A$ and $B$ with mass $m_A$ and $m_B$.

Around 100 seconds into the Big Bang, deuteron production peaks. This is followed by the production of helium, lithium, and beryllium. After approximately 3 minutes, the universe has cooled to the point where these nuclear reactions have ceased. Consequently, the abundances of the elements that formed during this time period persist to this day (with variations due to galactic chemical evolution).

In particular, the primary independent variables in the theory are the baryon-to-photon ratio, $\eta$, and the number of light neutrino species, \textit{N\textsubscript{$\nu$}}. Planck 2015 \cite{PlanckXIII} has provided more accurate observational constraints, yielding a value $(6.10\pm0.14)\times10$\textsuperscript{-10} for $\eta$. This fixed value allows the BBN theory to constrain alternative cosmologies\cite{Mathews17}.

In view of the the importance of BBN as a cosmological constraint, it is worthwhile to carefully analyze the uncertainties of the predicted BBN abundances. This is particularly true in the current age of precision cosmology, in which predicted abundances can be compared to observational results acquired from spectroscopy of metal-poor stars. Therefore, in the present work we undertake a Monte Carlo evaluation of the current uncertainties in the predictions of BBN.  This work is complementary to the Monte Carlo method employed in the contribution\cite{Coc17} by Coc and Vangioni to this volume as well as the analysis by Nakamura et al. \cite{Nakamura17} and the recent liklyhood analysis of Cyburt, Fields, and Olive.\cite{Cyburt16}

\section{\label{sec:level2}Methodology}

These updates included updated reaction rates from NACRE-II \cite{Nacre2}, REACLIB \cite{REACLIB}, and Descouvemont \cite{Descouvemont} for a myriad of reactions, including many of the 12 major reactions and auxiliary reactions involved in BBN. The major BBN reactions are given in Table 1. 


\begin{table}[ht]
\centering
\caption{Major BBN Reactions}
\begin{tabular}{|l||r|}
\hline
$p$ $\to$ $n$ & $t$ + $^{4}$He $\to$ $^{7}$Li + $\gamma$ \\ 

$p$ + $n$ $\to$ $d$ + $\gamma$ & $^{3}$He + $n$ $\to$ $t$ + $p$ \\

$d$ + $p$ $\to$ $^{3}$He + $\gamma$ & $^{3}$He + $d\to ^{4}$He + $p$ \\

$d + d \to ^{3}$He $ + n$ & $^{3}$He $ + ^{4}$He $\to ^{7}$Be + $\gamma$ \\

$d + d \to t + p$ & $^{7}$Li $ + p \to ^{4}$He $ + ^{4}$He \\

$t + d \to ^{4}$He $ + n$ & $^{7}$Be $ + n \to ^{7}$Li $ + p$ \\
\hline
\end{tabular}
    
\label{table:1}
\end{table}

We have incorporated these new reaction rates into a standard BBN code \cite{Kawano} which we have also checked with the more accurate Fermi integrals \cite{Fowler}. This code is available for public download \cite{KawanoCode}. We checked for variations in abundances when Bessel approximations in the code were replaced with these Fermi integrals. No significant changes in abundances were observed. 

NACRE provides reaction rates in tabular form featuring adopted values for a temperature $T_9$ (in units of $10^9$ K) (from 10 to 0.001) and its corresponding upper and lower bounds ($\sigma_u$ and $\sigma_l$). We consider the reaction rate uncertainties to have a gaussian distribution centered at the adopted value. Further, the dispersion for this gaussian was taken to be 
symmetric $\sigma = (\sigma_u - \sigma_l)/2$. A neutron lifetime of $880.3 \pm 1.1~s$ was also adopted. \cite{PDG}

After making these updates, we ran a Monte Carlo analysis. To do so, a Gaussian distribution of random numbers was created using the Box-Muller transform. This is a well-known method that converts two uniform distributions of random numbers on [0,1] into two normal distributed ones with a mean of 0 and $\sigma$ of 1 using the following equations:

\begin{equation}
        \Theta = 2\pi\textit{U\textsubscript{1}} \\
        \qquad \textit{R} = \sqrt{-2\ln{\textit{U\textsubscript{2}}}} \\
        \qquad \textit{X} = \textit{R}\cos{\Theta} \\
        \qquad \textit{Y} = \textit{R}\sin{\Theta} \\
\end{equation}

Here, $U_{1}$ and $U_{2}$ are uniform distributions of random numbers. They are converted into the transformation variables $R$ and $\Theta$. Then these variables yield two normal distributions of random numbers, indicated by $X$ and $Y$. As can be observed, this works even for a single uniformly distributed random number. Consequently, we took two uniformly distributed random numbers to create two normally distributed random numbers. We first took twelve of these normally distributed numbers. Each of these twelve random numbers is mapped to one of the 12 principal reaction rate gaussians. For a given $\eta$ value, these twelve random uncertainties are fixed in the beginning of the run. This is done because, for the most part, the S-factor varies little with energy. We have iterated this procedure 5000 times for each value of $\eta$ between $10^{-10}$ and $10^{-9}$. $2\sigma$ bounds were then determined. The resulting abundances and uncertainties as a function of $\eta$ are shown in Figure \ref{fig:1}.

\section{\label{sec:level3}Uncertainties}
The $\eta$ value of $(6.10\pm0.04)\times10$\textsuperscript{-10} determined by Planck is identified by the region with the black bars. For this $\eta$ value, the following 2$\sigma$ bounds are deduced and shown numerically in Eq. (4): 

\begin{figure}[ht]
\centering
\includegraphics[scale=.85]{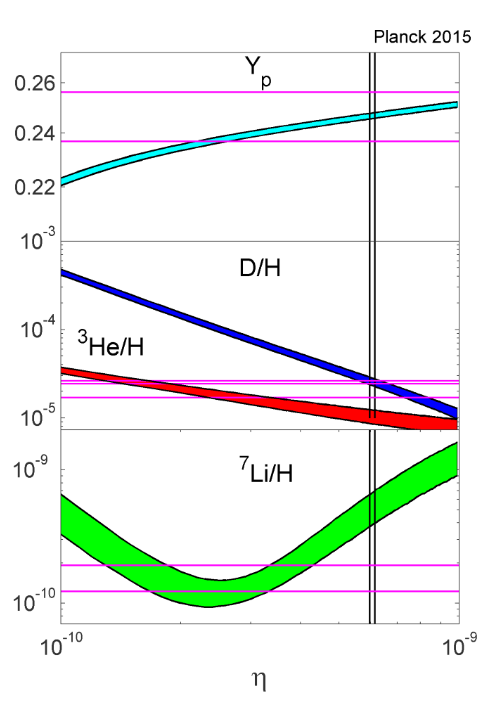}
\caption{Shaded bands show the 2$\sigma$ calculated abundance uncertainty as a function of  baryon-to-photon ratio $\eta$.  Horizontal lines indicated inferred primordial abundances from observation, while the vertical lines denote the value of $\eta$ deduced by the {\it Planck} analysis.}
\label{fig:1}
\end{figure}

\begin{equation}
\begin{gathered}
0.2456 < {Y_p} < 0.2478\\
2.2936\times10^{-5} < \mathrm{D}/\mathrm{H} < 2.7931\times10^{-5}\\
0.8520\times10^{-5} < ^3\mathrm{He}/\mathrm{H} < 1.2240\times10^{-5}\\
3.9344\times10^{-10} < ^7\mathrm{Li}/\mathrm{H} < 6.8499\times10^{-10}.\\
\end{gathered}
\end{equation}

The horizontal lines on Figure \ref{fig:1}  correspond to 2$\sigma$  bounds [Eq.~(5)] on observed light element abundances gathered from observational references summarized in Ref.~[\refcite{Cyburt16}], e.g. papers by Aver et al.,\cite{Aver} Pettini and Cooke,\cite{Pettini} Bania et al.,\cite{Bania} and Sbordone et al.,\cite{Sbordone} etc.].    (Note that 1$\sigma$ ranges are given in in Table 2):

\begin{equation}
\begin{gathered}
0.2369 < {Y_p} < 0.2529 \\
2.43\times10^{-5} < \mathrm{D}/\mathrm{H} < 2.63\times10^{-5} \\
0.90\times10^{-5} < ^3\mathrm{He}/\mathrm{H} < 1.30\times10^{-5} \\
1.00\times10^{-10} < ^7\mathrm{Li}/\mathrm{H} < 2.20\times10^{-10}.\\
\end{gathered}
\end{equation}

\begin{table}[ht]
\centering
\caption{BBN Abundances with 1$\sigma$}
\begin{tabular}{|p{1.3cm}||p{3.4cm}|p{3.4cm}|  }
\hline
Element&This paper&Observation Values\cr 
\hline
$Y\textsubscript{\textit{p}}$   &  $0.24670\pm.00056$   &$0.2449\pm0.0040$ \textsuperscript{[}\cite{Aver}\textsuperscript{]}\cr
 
D/H & $(2.54\pm0.12)\times10^{-5}$  & $(2.53\pm0.05)\times10^{-5}$ \textsuperscript{[}\cite{Pettini}\textsuperscript{]}\cr
 
$^3$He/H & $(1.038\pm0.093)\times10^{-5}$ & $(1.1\pm0.2)\times10^{-5}$ \textsuperscript{[}\cite{Bania}\textsuperscript{]}\cr
 
$^7$Li/H & $(5.39\pm0.73)\times10^{-10}$ &     $(1.6\pm0.3)\times10^{-10}$
\textsuperscript{[}\cite{Sbordone}\textsuperscript{]}\cr
\hline
\end{tabular}

\label{table:2}
\end{table}

As has been noted \cite{Cyburt16}, the theoretical predictions from BBN calculations agree quite well with observation. However, the calculated abundance for $^{7}$Li is about factor of 3 higher than the observational limits. This is known as the lithium problem, and it not surprising that this problem persists despite updated reaction rates and abundance uncertainties.

\begin{figure}[ht]
\centering
\includegraphics[scale=.485]{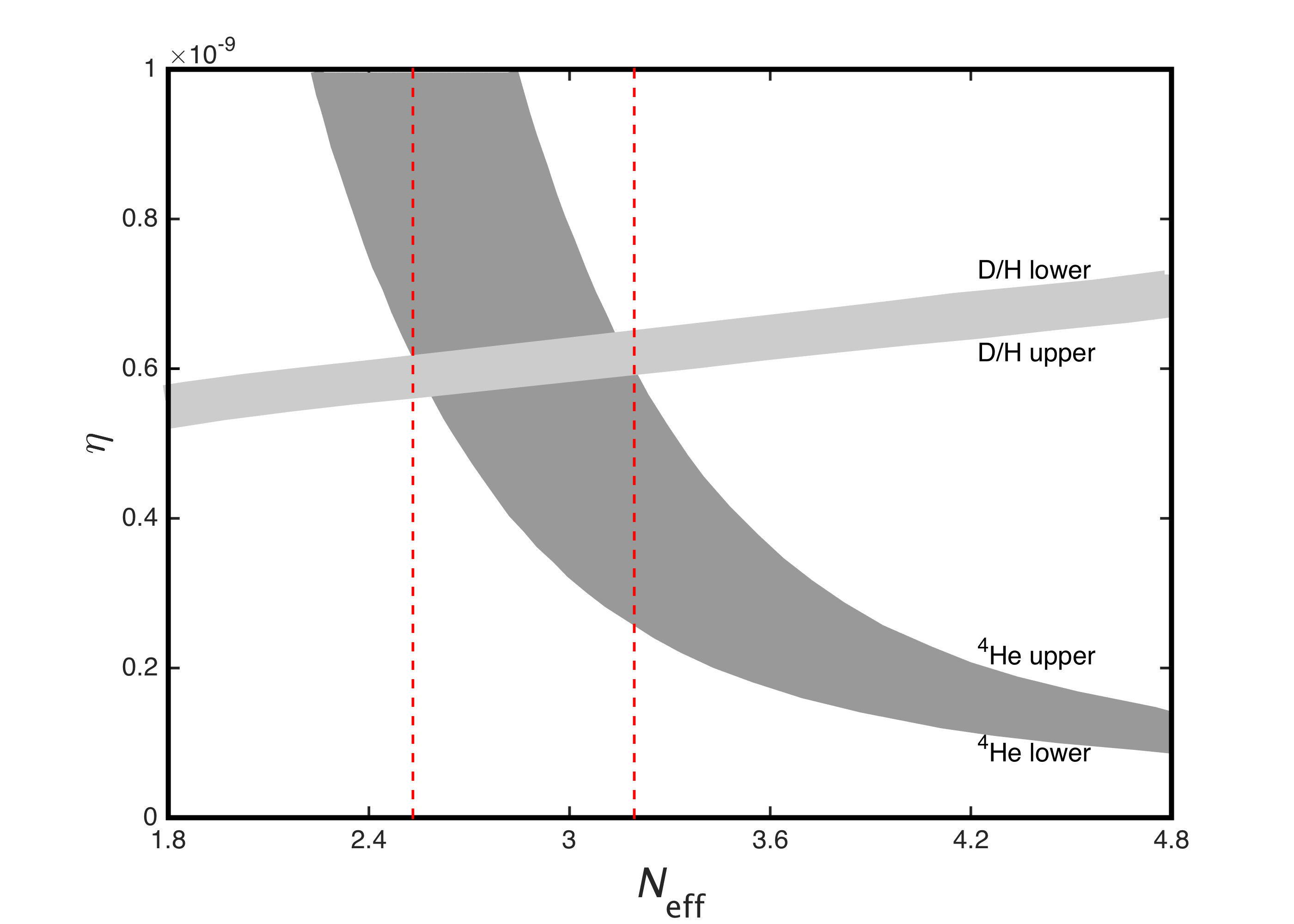}
\caption{1$\sigma$ bounds on $N_{eff}$. The bounds from BBN + observation occur at the intersections of $^4$He and D, which are shown by dotted vertical lines.}
\label{fig:2} \end{figure}

BBN can also be used to constrain the effective number of neutrinos, $N_{eff}$. Using our calculated uncertainties and the observational uncertainties cited above, we calculate new 1$\sigma$ bounds on $N_{eff}$. In Figure \ref{fig:2}, we derive constraints using only BBN and the observational uncertainties for $^4$He and D added in quadrature, giving bounds of $2.557 < N_{eff} < 3.174$. 

\section{\label{sec:level5}Conclusions}
Based on these new reaction rates and an updated neutron lifetime, we find a slight decrease in the calculational uncertainties  of elemental abundances when compared to previous work.  In order to solve the lithium problem, we reiterate that the solution may lie in new physics.\cite{Mathews17} These updated numbers can be used to study how new physics would affect primordial abundances in a more precise way. 

It also should be noted that this is a relatively simplistic analysis intended to demonstrate the effects of updated nuclear reaction rates and the Planck data on standard BBN. For a more thorough analysis or reaction rate uncertainties, please see Refs.~\refcite{Coc17,Nakamura17} in this volume as well as the recent review in Cyburt et al. \cite{Cyburt16}.

\section*{Acknowledgments}
Work at the University of Notre Dame was
 supported by the U.S. Department of Energy under Nuclear Theory Grant
 DE-FG02-95-ER40934.

\end{document}